%Paper: hep-th/9305123
%From: HUGHES@HUHEPL.HARVARD.EDU
%Date: Mon, 24 May 1993 14:20:25 -0400 (EDT)

\input harvmac
%\noblackbox
\Title{HUTP-93/A014}{\vbox{\centerline{Black Holes, Wormholes, and the
Disappearance}\vskip2pt\centerline{of Global Charge*}}}
\footnote{}{*Work supported in part by NSF Grant PHY92-18167}
\bigskip
\centerline{Sidney Coleman}
\medskip
\centerline{Shane Hughes}
\medskip
\centerline{Lyman Laboratory of Physics}
\centerline{Harvard University}
\centerline{Cambridge, MA 02138}
\bigskip
One of the paradoxes associated with the theory of the formation and
subsequent Hawking evaporation of a black hole is the disappearance of
conserved global charges.  It has long been known that metric fluctuations at
short distances (wormholes) violate global-charge conservation; if global
charges are apparently conserved at ordinary energies, it is only because
wormhole-induced global-charge-violating terms in the low-energy effective
Lagrangian are suppressed by large mass denominators.  However, such
suppressed interactions can become important at the high energy densities
inside a collapsing star.  We analyze this effect for a simple model of the
black-hole singularity. (Our analysis is totally independent of any detailed
theory of wormhole dynamics; in particular it does not depend on the wormhole
theory of the vanishing of the cosmological constant.)  We find that in
general all charge is extinguished before the infalling matter crosses the
singularity. No global charge appears in the outgoing Hawking radiation
because it has all gone down the wormholes.

\Date {05/93}

\newsec{Introduction}

There are two long-standing problems with Hawking radiation:  it contains too
much entropy and it contains too few baryons\ref\rHawk{S.~W.~ Hawking, Comm.
Math. Phys. {\bf43} (1975) 199; Phys. Rev. {\bf D14} (1976) 2460\semi
Ya.~B.~Zeldovich, Sov. Phys. JETP {\bf 45} (1977) 9.}.  The first is the
problem of the disappearance of quantum coherence during the formation and
subsequent evaporation of a black hole.  This is a very deep and difficult
problem and we shall say nothing about it here. The second is the problem of
the disappearance of conserved global charges during the same process. This
is the subject of this paper; we shall argue that this problem is neither
deep nor difficult.

Let us begin by stating the problem:  Consider the gravitational collapse of
a star with a nonzero value of some strictly conserved global charge.  For
example, the global charge could be $B-L$, and the star could be made of
helium.  In the late stages of collapse, the system begins to emit Hawking
radiation.  The Hawking radiation is formed outside the horizon and is totally
insensitive to the presence of nonzero $B-L$ in the collapsed star; it is
just as likely to contain quanta with one sign of $B-L$ as another. When the
star finally evaporates altogether, the initial $B-L$ has totally
disappeared, despite the fact that at every stage the dynamics of the process
has been governed by a Lagrangian that strictly conserves $B-L$.  \foot{This
problem can be avoided if the black hole does not evaporate altogether, but
instead leaves behind a Planck-scale remnant\ref\rRem{Y.~Aharonov, A.~Casher,
S.~Nussinov, Phys. Lett. {\bf B191} (1987) 51\semi T.~Banks, A.~Dabhokar,
M.~R.~Douglas and M.~O'Loughlin, Phys. Rev. {\bf D45} (1992) 3607.}.
However, the mechanism we shall present here works as well for
remnants, where it is not needed, as it does for total evaporation, where it
is.}

This problem doesn't arise for (continuous or discrete\ref\rKW{L.~Krauss
and F.~ Wilczek, Phys. Rev. Lett. {\bf 62} (1989) 1221.}) gauge charges.
These are detectable outside the horizon and can bias the Hawking radiation.
For example, when a virtual electron-positron pair appears outside a black
hole with positive electric charge, the black hole will attract the electron
and repel the positron; thus a positively charged black hole tends to produce
positively charged Hawking radiation.

Our proposed resolution of the problem rests on the fact that global charges
are distinguished from gauge charges in another way; they are the
charges that can disappear down a wormhole. A wormhole is a Euclidean field
configuration in some field theory containing gravity, consisting of two
asymptotically flat regions connected by a tube, or throat. Charge flowing
down a wormhole represents a quantum tunneling event in which charge
disappears from some small region of space-time and reappears in another
region, or even in a totally disconnected universe\ref\rWorm{For a review,
see A. Strominger, ``Baby Universes'', in {\it Particles, Strings and
Supernovae}, editors, A. Jevicki and C.-I. Tan, World Scientific, 1989.}. From
the viewpoint of a localized observer, the effect of the wormhole is a
violation of charge conservation.

But if this is the case, if, because of wormholes, there are no conserved
global charges, why are we tricked into believing  that $B-L$, for example,
is conserved?  The standard answer is that this is an accidental symmetry in
the sense of Weinberg.  The exact gauge symmetries of the theory are such
that all possible global-charge-violating interactions have mass dimension
$d>4$, and are therefore suppressed by factors of $M^{4-d}$, where $M$ is the
scale of the symmetry breaking.  (We work in units such that $\hbar=c=1$.)
The prototypical example of this is the original grand unified theory, the
$SU(5)$ model of Georgi and Glashow\ref\rGG{H. Georgi and S. Glashow, Phys.
Rev. Letters {\bf 32} (1974) 438.}.  In this theory, baryon-number-violating
processes occur on the grand unified scale, but the only
baryon-number-violating interactions one can construct have $d=6$; thus
baryon-number-violating amplitudes, like that for proton decay, are
suppressed by a factor of the inverse square of the grand unification mass.
In our case, the role of the grand unification mass is played by the wormhole
scale, the inverse size of the dominant wormholes in the functional integral.

These ideas lead to a novel picture of the fate of global charge during
gravitational collapse.  Let us imagine ourselves following infalling matter
through the horizon on its way to the singularity.  As the matter falls in,
its charge density and energy density become very large, and the
global-charge-violating interactions, negligible under ordinary conditions,
start to become strong.  When they are strong, they can drive the system
towards chemical equilibrium,  zero global-charge density.  Of course, at the
same time this is happening the matter is approaching the singularity.

Thus we have a race between equilibration and disaster.  If disaster wins, if
the matter reaches the singularity before the global-charge density is driven
to zero, we have to confront strongly-coupled quantum gravity to find the
fate of the global charge.  If equilibration wins (and we shall argue it
does) the matter streaming into the singularity carries no net global charge.
In this case, there is no problem of disappearing global charge.  No $B-L$ is
seen outside the black hole because there is no $B-L$ inside the black hole.
One way of phrasing things is to say that when the black hole evaporates
utterly, no $B-L$ is released because all the initial $B-L$ was cooked away
in the furnace of gravitational collapse.  Another, totally equivalent way of
phrasing things is to say no $B-L$ is released because all the initial $B-L$
has gone down wormholes\foot{This scenario is thus a realization of Hawking's
suggestion\ref\rHawktalk{S. Hawking, ``Black Holes and their Children, Baby
Universes'',  lecture given in Aspen Colorado, 5 July 1989.} that the missing
charge goes down a wormhole--although in this case it's not one big wormhole
but many little ones.  (Even earlier Zeldovich\rHawk  had suggested that the
missing charge ended up on a disconnected universe.)}.

Wormholes are a controversial subject, so we should stress that our arguments
do not depend in any substantial way on any detailed theory of wormhole
dynamics. In particular, they are completely independent of the much-disputed
wormhole theory of the cosmological constant\ref\rCosmo{S. Coleman, Nucl.
Phys. {\bf B310} (1988) 643\semi J. Polchinski Nucl. Phys. {\bf B324} (1989)
123; {\bf B325}
(1989) 619\semi W. Unruh, Phys. Rev. {\bf D40} (1989) 1053\semi  W. Fischler,
I.
Klebanov, J. Polchinski, and L. Susskind, Nucl. Phys. {\bf B327} (1989) 157.}.
All we'll need for our work is the statement that global-charge-violating
interactions become strong at some large mass scale, $M$.  For energy $E\ll
M$, the leading effects of these interactions are represented by an operator
of high ($>4$) dimension, that is to say, by a charge-violating Lagrange
density of the form
\eqn\eIi{{\cal L}_{CV}=M^{4-d}\times({\rm operator\ of\ order\ one}).}

The most natural choice for $M$ would be the Planck mass, $M_P$.  However, for
most of our work we'll choose $M$ to be a few orders of magnitude less than
$M_P$.  This is to keep ourselves honest; it keeps the physics we're
interested in well away from the regime of strongly-coupled quantum gravity.
Most of our conclusions would formally go through even if $M$ were on the
order of $M_P$, but our approximations would be totally unreliable, because
the effects we compute would be of the same order of magnitude as the effects
of gravitational fluctuations which we neglect.\foot{For what it's worth, in
the wormhole theory of the cosmological constant it is possible to arrange
matters such that the wormhole scale is indeed a few orders of magnitude less
than $M_P$, without any fine-tuning of parameters\ref\rCL{S. Coleman and
K. Lee, Phys. Lett. {\bf B221} (1989) 242.}.}

To determine whether equilibration or disaster prevails, we must analyze
the rate equations for the disappearance of global charge. In Sec. 2 we
perform such an analysis for a very simple model of the singularity at
the core of the collapsing star, a collapsing radiation-filled
Friedman-Robertson-Walker universe.  This model is inspired by the
Oppenheimer-Snider solution for the gravitational collapse of a spherically
symmetric dust cloud.  Here the inside of the collapsing cloud is a section
of a dust-filled collapsing Friedman-Robertson-Walker universe.  We replace
Oppenheimer and Snider's dust by radiation (that is to say, by
ultrarelativistic matter) because the physics we are interested in takes
place on energy scales much larger than the relevant particle masses. We
don't claim that this is in any sense a good approximation to the generic
black-hole singularity, or even the interior part of some
Oppenheimer-Snider-like solution, but its homogeneity and isotropy makes it
especially easy to analyze, and thus a good place to begin an investigation.

The result of our analysis is surprisingly simple:  under the stated
conditions, equilibration always wins.  The disappearance of global charge
need have nothing to do with strongly-coupled quantum gravity.

Sec. 3 discusses both possible alternatives to wormholes and the question of
to what extent we expect our results to survive if we weaken our assumptions.
The analysis is still in progress, but, at the moment, things look good.

\newsec{Rate Equations in a Collapsing Universe}
\subsec{Preliminaries}

The Einstein equation for a Robertson-Walker metric is
\eqn\eIIi{ {\dot R^2\over R^2}={{8\pi}\over{3M_P^2}}\rho -{k\over R^2}.}
where $R$ is the scale factor, $\rho$ the energy density, and $k=0, \pm 1$.
As always, this must be supplemented by an equation of state, giving the
pressure as a function of the energy density.  As explained in Sec. 1, we
take the equation of state to be that of an ultrarelativistic gas,
$p=\rho/3$. Then $R$, for small $t$, is proportional to $\sqrt{|t|}$, where
we have chosen the zero of time so the singularity, $R=0$, occurs at $t=0$.

It will be convenient for us to define an ``energy length'', $\ell$, by
\eqn\eIIii{\rho=\ell^{-4}.}
Then, from the Einstein equation, for small $t$,
\eqn\eIIiii{\ell\sim\sqrt{{{|t|}\over{M_P}}},}
where $\sim$ denotes equality up to numerical constants.  This is the
only equation of general relativity we shall need in our work.

To keep things as simple as possible, we'll speak in this section as if we
were studying only one global charge, and as if the total charge was the
difference between the number of particles (objects with global charge one)
and antiparticles (objects with global charge minus one).  The extension to
many global charges and many types of particles carrying various amounts of
the several charges is trivial.

We shall follow the time evolution of
\eqn\eIIiv{\hat n=j_0\ell^3,}
where $j_\mu$ is the current of the global charge.  Because $\ell/R$ is
a constant of the motion, this is equivalent to following the global-charge
density per unit comoving volume.

We'll assume throughout our computations that our ultrarelativistic gas is
locally in thermal equilibrium (although not of course in chemical
equilibrium), so its local state is completely determined by $\hat n$ and
$\ell$ (equivalently, by the temperature and the chemical potential). Even
when $\dot R/R$ is large, this is a safe assumption; for an
ultrarelativistic gas, a state of thermal equilibrium stays a state of
thermal equilibrium (with a blue-shifted temperature) under uniform
contraction of the universe.

\subsec{Lowest-Order Perturbation Theory in Both ${\cal L}_{CV}$ and $\hat n$}

The easiest situation to analyze is that in which the interaction is weak (so
we can treat the rate of charge disappearance to lowest (second) order in
the charge-violating interaction) and the difference between the number of
particles and the number of antiparticles is small compared to their sum
(so we can linearize the rate equation in $\hat n$).  The rate equation is
then determined completely, up to numerical multiplicative constants:
\eqn\eIIv{{{d\hat n}\over{dt}}\sim -{{\hat n}\over{\ell(M^2\ell^2)^{d-4}}}.}
The power of $M$ in this equation is determined by the fact that it is
second order in ${\cal L}_{CV}$; the power of $\ell$ is then fixed by
dimensional analysis.  $M_P$ does not appear because this is a purely
microphysical equation, computed (in principal) by applying kinetic theory
in locally inertial coordinates.

Since $\ell$ is a proportional to a power of $t$, integrating this with
respect to $t$ is (up to numerical constants), the same as multiplying by
$t$.  Thus,
\eqn\eIIvi{\ln\hat n\sim -{|t|\over{\ell(M^2\ell^2)^{d-4}}}+{\rm constant}}
Using Eq. \eIIiii , we find
\eqn\eIIvii{\lim_{t\to0}{\hat n}=0 \qquad{\rm for\ } d>4\half .}
In particular, the global charge is always totally extinguished, vanishing
more rapidly than any power of $t$, for the interactions that interest us,
$d\geq 5$.

Of course, these formulas are not to be trusted for very small $t$, when the
interaction becomes strong and lowest-order perturbation theory is
inapplicable.  We'll take care of this regime shortly; meanwhile, it's
interesting to ask how much of the global charge has disappeared by the time
the interaction does become strong. At $\ell M \sim 1$,
\eqn\eIIviii{{|t|\over{(M^2\ell^2)^{d-4}\ell}}\sim{M_P\over M}.}
Thus only $\exp(-O(M_P/M))$ of the original charge remains.  If (as we have
been assuming) $M$ is a few orders of magnitude less than $M_P$, this is
not much charge at all.

\subsec{Lowest-Order Perturbation Theory in ${\cal L}_{CV}$ but not in $\hat
n$}

Up to now we have assumed that the difference between the density of
particles and of antiparticles was small compared to their sum. This
assumption is certainly not true in the early stages of collapse, when the
star contains only particles, and it is not obvious whether sufficient
antiparticles to make it true are produced later on.  Therefore in this
subsection we eliminate this assumption.

We still assume that the charge-violating interaction is weak, so we can use
second-order perturbation theory in ${\cal L}_{CV}$.  Eq. \eIIv\ is replaced
by
\eqn\eIIix{{{d\hat n}\over{dt}}=-{{\hat nr(\hat
n)}\over{(M^2\ell^2)^{d-4}\ell}},}
where $r(\hat n)$ is some dimensionless function of $\hat n$.  We know from
the previous analysis that $r$ is positive at zero.  Because there is no
equilibrium state other than $\hat n=0$, $r$ must be positive everywhere.

Integrating this equation from some initial time $t_0$, we find
\eqn\eIIx{\int_{\hat n(t_0)}^{\hat n(t)} {{d \hat n}\over{\hat nr(\hat n)}}
=-\int_{t_0}^t{{dt}\over{(M^2\ell^2)^{d-4}\ell}}.}
For the cases of interest to us ($d \geq 5$), the right-hand side of this
equation blows up as $t$ goes to zero.  Therefore, so must the left-hand
side.  But since $r$ is everywhere positive, this can only happen if $\hat n$
goes to zero.

We have computed $r$ numerically and integrated Eq. \eIIx\ for a simple case,
the theory of a single massless Dirac field with particle-number-violating
interaction
\eqn\eIii{{\cal L}_{CV}=M^{-2}(\bar\psi^c\psi)^2,}
where $\psi^c$ is the charge conjugate of $\psi$.  In lowest-order
perturbation theory, all the partial-wave amplitudes for this theory are
proportional to $s$, the square of the center-of-mass energy.  An easy
calculation shows that the largest of these hits the unitarity bound
at $s\approx 5M^2$.  This gives us a rough measure of strong coupling:  as
long as the average value of $s$ is well below $5M^2$, we are well clear of
the strong-coupling regime.

In our numerical computation, we chose the initial condition $\hat
n(-\infty)=\sqrt{8/9\pi}$, the appropriate value for a cold gas of massless
fermions with no antifermions.  Our result is shown in the figure; we have
chosen coordinates such that the linear (small $\hat n$) approximation (shown
by the dashed line) is a straight line.  The graph is indistinguishable
from the linear approximation for the average value of $s/5M^2$  greater than
$6(M/M_P)^{2/3}$; for $M$ a few orders of magnitude below $M_P$, this is well
in the weak-coupling region.   Note that even for earlier times, the linear
approximation is a conservative estimate, in that it underestimates the
rate of charge extinction.

\subsec{The Strong Coupling Regime}

We now have a good picture of what happens to global charge from the
beginning of collapse to the time when $\ell M$ approaches one; at the end of
this epoch, most of the global charge has been cooked away, rate equations
linear in $\hat n$ are a good approximation, and the charge-violating
interaction is about to become strong.  We can not go beyond this point with
dynamical calculations, because we can only compute wormhole effects in
weak-coupling approximations. Instead, we will resort to the old idea that
when interactions become strong, cross sections become as large as they can
be, consistent with the finite range of the interactions. This leads to
constant cross-sections at high energies, possibly modified by slow
(logarithmic) growth.  This is in excellent agreement with experiment for the
one strongly-interacting relativistic theory on which experiments have been
performed, hadrodynamics.

Thus we assume that for energies greater than $M$, the total cross-section
for charge-violating scatterings is
\eqn\eIIxi{\sigma_{CV}\sim M^{-2},}
possibly times logarithms of the energy, which will not affect our argument.
Here we've guessed that the range of the interactions is on the order of
$1/M$, the only relevant length in the problem. If we underestimate the rate
of extinction of global charge by only including the effects of two-body
scatterings, we find, on the usual dimensional grounds, that
\eqn\eIIxii{{{d\hat n}\over{dt}}\sim -{{\hat n}\over{M^2\ell^3}}.}
This is identical to Eq. \eIIv , second-order perturbation theory, for $d=5$,
(although, of course, the angular distribution of the outgoing particles is
completely different), and, like it, it leads to total extinction of global
charge.

\newsec{Comments}

We have found that, if there are wormhole-induced charge-violating
interactions that are not observable at low energies because of the presence
of inverse powers of the wormhole scale, then these interactions {\it always}
remove all the global charge from a collapsing star, for a certain model
of the collapse singularity.  The obvious questions are:  To what extent
does our analysis depend on wormholes?  To what extent does our analysis
depend on our model of the singularity?

We can think of three mechanisms that might substitute for wormholes:

(1)  Otherwise unspecified metric fluctuations at the Planck scale:
All of the equations of Sec. 2 apply in this case, with $M$ set equal to
$M_P$, and they lead to the same conclusion, the total extinction of the
global charge.  The problem here, as we remarked in the Introduction, is
that in addition to the physics described in Sec. 2, there is the physics
of strongly-coupled quantum gravity, which should be important at this scale,
and which is totally ignored in Sec. 2, for the excellent reason that we
don't know anything about it.  To phrase the same thought in another way,
it's silly to track global charge when the infalling matter is a Planck
time away from the singularity; when you're a Planck time away from the
singularity, fluctuations in the metric are so large that concepts like
``time'' and ``singularity'' lose their meaning.  Of course it's always
possible that someday, when strong quantum gravity effects can be treated
reliably, it will turn out that the arguments of Sec. 2 are basically
correct.  However, as H. G. Wells said in another context, ``If anything
is possible, nothing is interesting.''

(2)  Strings: It is essentially impossible to implement a continuous symmetry
in string theory that is not a gauge symmetry.  Thus strings are as much
enemies of conserved global charges as wormholes.  From this point on, the
discussion is the same as that of the preceding paragraph.

(3)  Grand unified theories:  The absence of nongauged continuous symmetries
in grand unified theories is as much a matter of fashion as of physics;
nevertheless, grand unified theories do have one of the key feature of the
theories we have been discussing, charge-violation suppressed by large mass
denominators, and it is worth asking to what extent our analysis applies to
them.   The big difference is that the grand unified interactions never
become strong; at the grand unification scale, $M_{GUT}$, they are weak (and
renormalizable), and they stay weak at higher energies.  This has only a
mild effect on the arguments of Secs. 2.2 and 2.3.  For example, in Eq.
\eIIvi , the factor of $(1/M^2)^{d-4}$ is replaced by a factor of
$\alpha_{GUT} (1/M_{GUT}^2)^{d-4}$ (assuming the underlying process is first
order in the grand unified gauge coupling).  This does not make a big
difference; for reasonable choices of the parameters, most of the global
charge is still extinguished.  However, the last part of the extinction,
discussed in Sec. 2.4, does not work; the mild high-energy behaviour allows
the remaining global charge to reach the singularity.

As for the question of the singularity, there is a clear path to take.
The mixmaster singularity\ref\rMix{C.~W.~Misner, K.~S.~Thorne and
J.~A.~Wheeler, {\it Gravitation}, W.~H.~Freeman, 1973, pp 805-814;
L.~D.~Landau and E.~M.~Lifshitz, {\it The Classical Theory of Fields},
Pergamon Press, 1975, pp 390-397.} is the outstanding candidate for a generic
black-hole singularity.  The mixmaster is neither isotropic nor homogeneous,
and does not preserve thermal equilibrium; thus it presents a more difficult
problem than the cosmological singularity, but it is not beyond analysis. We
have successfully analyzed the fundamental component of the mixmaster, the
homogeneous Kasner universe, and have found, once again, total extinction of
global charge.  (Although the rate of extinction is quite different from the
cosmological case; at the end, $\hat n$ only vanishes like a power of $t$.)
We hope to extend our analysis to the full mixmaster and report on our
results shortly.

One of us (S.C.) is grateful for conversations with Charles Alcock, Andrew
Cohen, Howard Georgi, Roger Penrose, and Edward Witten.
\vskip2in
\centerline {\bf Figure Caption}
\bigskip
\noindent The result of a numerical computation, for ${\cal
L}_{CV}=M^{-2}(\bar\psi^c\psi)^2$, of the charge per comoving volume in a
collapsing universe.  The system was started as a cold gas of particles (with
no antiparticles) at $t=-\infty$.  The dashed line is the linear
approximation to the rate equation.

\bigskip
\listrefs
\bye